\documentclass[aps,twocolumn,english,superscriptaddress,citeautoscript,showkeys,preprintnumbers,amsmath,amssymb,floatfix,footinbib,prl]{revtex4-2}

\usepackage{graphicx}
\usepackage{xcolor}
\usepackage{soul}
\usepackage{amssymb,amsmath}
\usepackage{hyperref}
\hypersetup{colorlinks=true,linkcolor=red,citecolor=blue}
\usepackage[utf8]{inputenc}
\usepackage[english]{babel}
\usepackage{booktabs}
\usepackage{comment}

\newcommand{\Tc}{$T_\text{c}$}
\begin{document}

\title{Vacancy-free cubic superconducting NbN enabled by quantum anharmonicity}

\author{Eva Kogler}
\affiliation{Institute of Theoretical and Computational Physics, Graz University of Technology, NAWI Graz, 8010, Graz, Austria}

\author{Mihir R. Sahoo}
\affiliation{Institute of Theoretical and Computational Physics, Graz University of Technology, NAWI Graz, 8010, Graz, Austria}

\author{Chia-Nien Tsai}
\affiliation{Smead Department of Aerospace Engineering Sciences, University of Colorado, Boulder, Colorado 80303}

\author{Fabian Jöbstl}
\affiliation{Institute of Theoretical and Computational Physics, Graz University of Technology, NAWI Graz, 8010, Graz, Austria}

\author{Roman Lucrezi}
\affiliation{Institute of Theoretical and Computational Physics, Graz University of Technology, NAWI Graz, 8010, Graz, Austria}
\affiliation{Department of Materials and Environmental Chemistry, Stockholm University, SE-10691 Stockholm, Sweden}

\author{Peter I. C. Cooke}
\affiliation{Department of Materials Science and Metallurgy, University of Cambridge, 27 Charles Babbage Road, Cambridge, CB3 0FS, UK}

\author{Birgit Kunert}
\affiliation{Institute of Solid State Physics, Graz University of Technology, NAWI Graz, 8010, Graz, Austria}

\author{Roland Resel}
\affiliation{Institute of Solid State Physics, Graz University of Technology, NAWI Graz, 8010, Graz, Austria}

\author{Chris J. Pickard}
\affiliation{Department of Materials Science and Metallurgy, University of Cambridge, 27 Charles Babbage Road, Cambridge, CB3 0FS, UK}
\affiliation{Advanced Institute for Materials Research, Tohoku University, Sendai, 980-8577, Japan}

\author{Matthew N. Julian}
\affiliation{Intellectual Ventures, Bellevue, Washington, United States}

\author{Rohit P. Prasankumar}
\affiliation{Intellectual Ventures, Bellevue, Washington, United States}

\author{Mahmoud I. Hussein}
\affiliation{Smead Department of Aerospace Engineering Sciences, University of Colorado, Boulder, Colorado 80303}
\affiliation{Department of Physics, University of Colorado, Boulder, Colorado 80302}

\author{Christoph Heil}
\email[Corresponding author: ]{christoph.heil@tugraz.at}
\affiliation{Institute of Theoretical and Computational Physics, Graz University of Technology, NAWI Graz, 8010, Graz, Austria}

\begin{abstract}
Niobium nitride (NbN) is renowned for its exceptional mechanical, electronic, magnetic, and superconducting properties. The ideal 1:1 stoichiometric $\delta$-NbN cubic phase, however, is known to be dynamically unstable, and repeated experimental observations have indicated that vacancies are necessary for its stabilization. In this work, we demonstrate that when the structure is fully relaxed and allowed to distort under quantum anharmonic effects, a previously unreported stable cubic phase with space group $P\bar{4}3m$ emerges --- 65\,meV/atom lower in free energy than the ideal $\delta$ phase. This discovery is enabled by state-of-the-art first-principles calculations accelerated by machine-learned interatomic potentials. To evaluate the vibrational and superconducting properties with quantum anharmonic effects accounted for, we use the stochastic self-consistent harmonic approximation (SSCHA) and molecular dynamics spectral energy density (SED) methods. Electron-phonon coupling calculations based on the SSCHA phonon dispersion yield a superconducting transition temperature of \mbox{\Tc{} = 20\,K}, which aligns closely with experimentally reported values for near-stoichiometric NbN. These findings challenge the long-held assumption that vacancies are essential for stabilizing cubic NbN and point to the potential of synthesizing the ideal 1:1 stoichiometric phase as a route to achieving enhanced superconducting performance in this technologically significant material.
\end{abstract}

\date{\today}


\maketitle

\section*{Introduction} 
Transition metal carbides and nitrides combine exceptional mechanical properties with diverse electronic, magnetic and superconducting behavior~\cite{Chen-2005, Chen-2-2005, Soignard-2007, WILLIAMS1967333, Kasumov-1999, Zou2015}. Among these materials, niobium nitride (NbN) has attracted significant attention due to its structural complexity and technological relevance. NbN exhibits a rich phase diagram due to its ability to accommodate varying nitrogen concentrations and vacancy contents. Typically, the nitrogen content is quantified by \(x\) in the formula \(\mathrm{NbN}_x\), representing the ratio of nitrogen to niobium. With increasing \(x\), a variety of phases has been reported: the \emph{bcc} solid solution \(\alpha\)-NbN (\(\mathrm{NbN}_x\) with \(x < 0.40\), $Im\bar{3}m$), the hexagonal \(\beta\)-NbN ($P\bar{6}m2$, \(x = 0.40\)–0.50), the tetragonal \(\gamma\)-NbN ($Cmcm$, \(x = 0.75\)–0.80), the hexagonal \(\eta\)-NbN ($P6_3/mmc$, \(x = 0.80\)–0.95), the \emph{fcc} \(\delta\)-NbN ($Fm\bar{3}m$, \(x = 0.88\)–0.98 and \(x = 1.015\)–1.062), and the hexagonal \(\epsilon\)-NbN ($P6_3/mmc$, \(x = 0.92\)–1.00). Additional phases such as tetragonal \(\mathrm{Nb}_4\mathrm{N}_5\), hexagonal \(\delta'\)-NbN ($P6_3/mmc$, \(x = 0.93\)–0.98), and hexagonal \(\epsilon'\)-\(\mathrm{NbN}\) ($P6_3/mmc$, at \(x = 1\)) have also been observed.

Although $\epsilon$-NbN is predicted to be the ground-state structure at $x = 1$ from first-principle calculations, $\delta$-NbN is the phase most commonly observed under ambient-pressure conditions. It also exhibits the highest superconducting transition temperature ($T_\text{c}$), reaching up to 17\,K. Other superconducting phases include $\gamma$-NbN and Nb\textsubscript{4}N\textsubscript{5}, though with lower $T_\text{c}$ values~\cite{Oya-1974, Aschermann1941, Keskar_1971}.
Despite the emergence of superconducting materials with higher \Tc~\cite{shen2008cuprate,kresin2013fundamentals,hosono2015iron,Boeri_JPCM_2021_roadmap,wang2024experimental}, NbN remains a subject of intense study because of its high critical magnetic field, robust chemical and mechanical stability, and ease of synthesis and processing~\cite{Cucciniello2022}. These attributes have led to applications in superconducting radiofrequency cavities~\cite{BENVENUTI199316}, hot electron photodetectors~\cite{LINDGREN1998423}, Josephson junctions~\cite{Setzu2008}, qubits~\cite{Nakamura-2011}, and superconducting nanowire single-photon detectors~\cite{marsili2012efficient,simon2025ab}.

Typically, $\delta$-NbN is synthesized with a random distribution of nitrogen vacancies, which induce structural distortions that stabilize the $\delta$-phase. As vacancy concentration decreases, these distortions diminish, resulting in a higher \Tc{} until reaching a critical threshold where the structure becomes unstable and transforms into the $\epsilon$-NbN phase~\cite{Oya-1974}. This interplay between vacancy-induced distortions, structural stability, and superconducting properties highlights the importance of precise stoichiometric control.

\emph{In silico} simulation of disorder and vacancies is computationally demanding because it necessitates the use of very large supercells in addition to statistical models~\cite{zunger1990,wei1990,ferreira2024}, thus, theoretical research focus has been mostly on the ideal 1:1 stoichiometry, where $\delta$-NbN is dynamically unstable within the harmonic approximation. Previous work using the Korringa-Kohn-Rostoker Coherent Potential Approximation revealed that the presence of vacancies lowers the Fermi level and increases band smearing around it~\cite{Marksteiner-1986}. Building on this observation, applying a large electronic smearing --- which can also serve as a first-order approximation for temperature effects --- has been shown to eliminate the dynamic instability~\cite{Ivashchenko2010, Babu-2019}. Similarly, within the virtual crystal approximation, the stabilization of cubic NbN has been achieved, yielding phononic properties that compare favorably with experimental data, albeit with an overestimated electron-phonon coupling~\cite{Babu-2019}.

In this study, we reexamine the $\delta$-NbN structure and uncover a previously unreported stable phase at 1:1 stoichiometry. We first demonstrate that the conventional $\delta$ structure remains unstable, even when quantum anharmonic effects are taken into account---specifically, the combined influence of quantum ionic motion (such as zero-point motion, absent in classical treatments) and the non-parabolic nature of the ionic potential---using the stochastic self-consistent harmonic approximation (SSCHA)~\cite{Werthamer_1970,Monacelli_2021}, and considering temperatures up to 300\,K. By exploring the potential energy surface beyond the Born-Oppenheimer approximation, explicitly accounting for these quantum anharmonic effects, we identify a distinctly stable, displaced structure within the simple cubic cell. Our results from two complementary computational approaches --- SSCHA and molecular-dynamics spectral energy density (MD-SED) calculations~\cite{larkin2014comparison,honarvar2016spectral}---show excellent agreement, and are further supported by an independent \emph{ab initio} random structure search (AIRSS)~\cite{pickard2006high,AIRSS}.
The newly identified structure features a reduced electronic density of states at the Fermi level and markedly weakened electron-phonon coupling due to its lattice distortion. Moreover, the superconducting transition temperature (\Tc), calculated using the isotropic Migdal-Eliashberg (ME) formalism, closely matches experimental values for bulk NbN close to the 1:1 stoichiometry.

These findings highlight the critical role of quantum anharmonicity, i.e. quantum ionic motion, phonon-phonon interactions, and the non-parabolic shape of the ionic potential, in stabilizing a new simple cubic phase of NbN --- one that is thermodynamically more favorable than the $\delta$ phase, though still higher in energy than the $\epsilon$ phase. Beyond this insight, our results point to a promising new experimental direction. Whereas previous efforts have largely focused on tuning vacancy concentrations to tailor material properties, our findings suggest that synthesizing the ideal 1:1 stoichiometric cubic phase could offer a more direct path to enhancing superconducting performance.

\section*{Results and Discussion}
\begin{figure*}[t]
    \centering
    \includegraphics[width=1\linewidth]{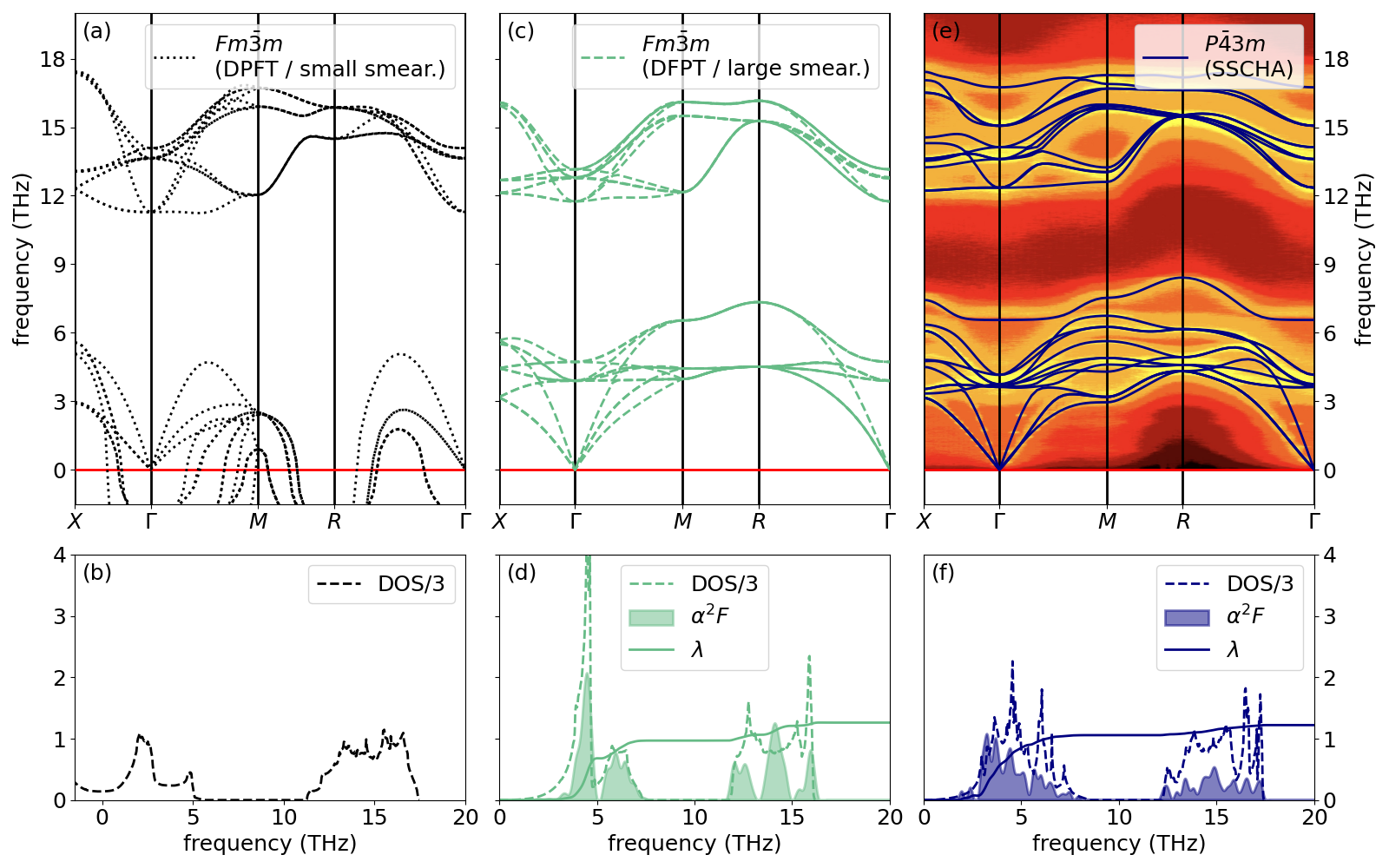}
    \caption{Top row: (a) phonon dispersion of $Fm\bar{3}m$ in the harmonic approximation (dotted black), (c) of $Fm\bar{3}m$ in the harmonic approximation using a large smearing parameter (dashed green), and (e) $P\bar{4}3m$ structure within SSCHA at 15\,K (solid blue). The colored background in (e) shows the spectral density obtained with MD-SED.
    Bottom row: phonon DOS (dashed line, rescaled by a factor of 1/3 to enable same scales with the other quantities), and $\alpha^2 F$ (filled line) and cumulative $\lambda$ (solid line) for (b) $Fm\bar{3}m$ in the harmonic approximation, (d) $Fm\bar{3}m$ in the harmonic approximation using a large smearing parameter, and (f) $P\bar{4}3m$ structure within SSCHA at 15\,K.}
    \label{fig:comp_phonons}
\end{figure*}
We begin our discussion by examining the phonon and electron-phonon properties of $\delta$-NbN within the harmonic approximation. For the ideal 1:1 stoichiometry, several imaginary modes appear over large portions of the Brillouin zone, as shown in Fig.~\ref{fig:comp_phonons}(a). This indicates that the structure is dynamically unstable, in agreement with experimental observations that vacancies must be introduced to stabilize the $\delta$ phase. Modeling a structure that includes vacancies and the resulting disorder is extremely challenging because it requires the use of exceptionally large supercells, a task that quickly becomes computationally prohibitive due to the $\mathcal{O}(N^3)$ -- $\mathcal{O}(N^4)$ scaling of current phonon and electron-phonon calculation methods based on plane-wave DFT, where $N$ is the number of electrons in the computational cell~\cite{giustino2014materials,giustino2017electron}. Moreover, the prevalence of imaginary phonon frequencies throughout the Brillouin zone precludes reliable calculations of superconducting properties for the ideal structure.

A commonly employed, computationally efficient workaround is to use an unusually large electronic smearing parameter in DF(P)T calculations. This approach effectively averages the Fermi surface and related properties, thereby yielding a harmonically stable phonon dispersion, as demonstrated in Fig.~\ref{fig:comp_phonons}(c), and showing good agreement with neutron diffraction data on NbN~\cite{christensen_1979}. However, when the electron-phonon interaction is evaluated under this approximation and superconducting properties are subsequently calculated (see Tab.~\ref{tab:tc}), the resulting critical temperature is significantly overestimated (27\,K compared to the experimental value of about 16\,K). This discrepancy leads us to conclude that the large smearing approximation does not adequately capture the true superconducting behavior of the system.

To better discern the shortcomings stemming from the harmonic approximation versus those arising from treating NbN as a perfect 1:1 crystal, we employed two complementary methods to evaluate its vibrational properties: the SSCHA and MD-SED. In the SSCHA framework, quantum ionic motion and anharmonic effects are explicitly incorporated to generate an effective harmonic description that more accurately reflects the true behavior of the system. In contrast, the MD-SED approach involves full molecular dynamics simulations to compute the spectral functions, thereby inherently capturing ionic motion and anharmonicity to all orders. More details on these methods and further references can be found in the Methods section and Supplementary Information (SI)~\cite{SI}.

Because both SSCHA and MD-SED require performing hundreds of thousands of DFT calculations for full convergence, carrying out these computations solely at the DFT level is not feasible. Instead, we used DFT-calculated data to train high-fidelity machine-learned \textit{ab initio} interatomic potentials that retain DFT-level accuracy in energies and forces. This approach enabled us to converge the SSCHA and MD-SED calculations with a computational speed-up of approximately $10^4$--$10^5$~\cite{lucrezi2023quantum}. Detailed information on the construction of these potentials and their validation benchmarks is provided in the Methods section and SI~\cite{SI}.

Employing the SSCHA on the $\delta$-NbN phase, we still observed imaginary phonon frequencies even at temperatures up to 300\,K~\footnote{SSCHA calculations at 0\,K reveal a lattice instability through the presence of imaginary phonon frequencies, regardless of the supercell size. At 300\,K, calculations using a 2$\times$2$\times$2 supercell of the primitive \emph{fcc} unit cell yield fully positive phonon frequencies. However, larger supercells at the same temperature again show imaginary modes, indicating that the smallest supercell size gives unconverged results.}. This indicates that the ideal 1:1 stoichiometric $\delta$ phase is dynamically unstable --- even when quantum ionic motion and anharmonic effects are incorporated --- which is consistent with experimental findings that vacancies must be introduced to stabilize this phase.

\begin{figure}[t]
    \centering
    \includegraphics[width=1\linewidth]{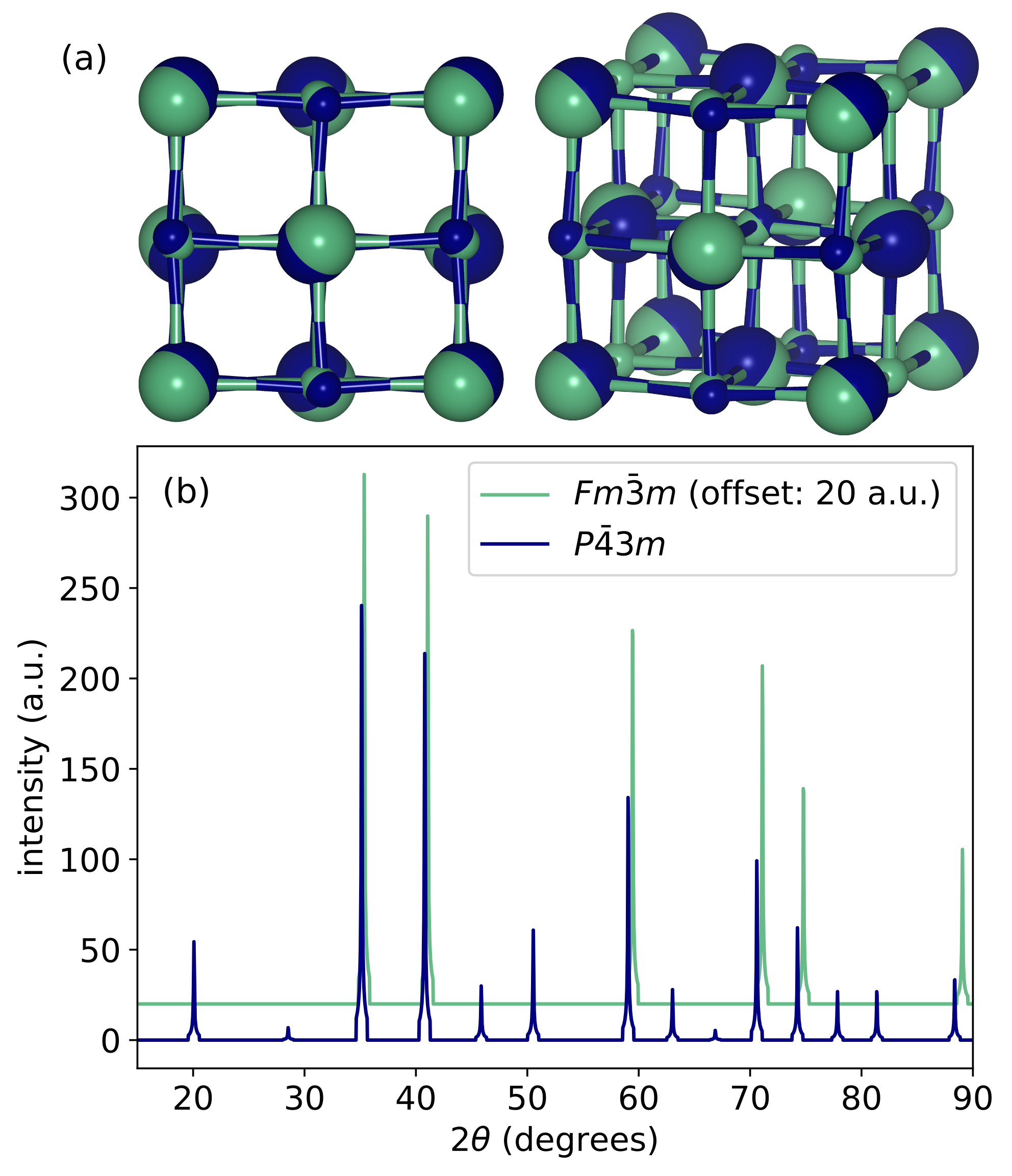}
    \caption{(a) Superposition of the $Fm\bar{3}m$ (green) and $P\bar{4}3m$ (blue) NbN crystal structures, showing all atoms of each structure in a single color. The overlay highlights how $P\bar{4}3m$ distorts from the high-symmetry $Fm\bar{3}m$ phase. Structures visualized with VESTA~\cite{Momma-vesta-2011}. (b) Simulated XRD patterns using Cu K$\alpha$ radiation~\cite{Kraus1998}.}
    \label{fig:comp_phases}
\end{figure}

Next, we allowed the entire structure to fully relax within the SSCHA framework, effectively permitting the system to follow the soft phonon mode displacements~\footnote{Closer inspection of the structure obtained from the full SSCHA relaxation shows that its atomic arrangement can be described as a linear combination of the eigenvectors of a threefold-degenerate $\Gamma$ mode with $T_{1u}$ symmetry in the conventional (primitive cubic) unit cell of the $Fm\bar{3}m$ structure. In the corresponding harmonic phonon calculation [Fig.~\ref{fig:comp_phonons}(a)], this mode exhibits the largest imaginary frequency of approximately $7.4 i$\,THz at $\Gamma$.}. This procedure led to the emergence of a novel cubic NbN phase with space group $P\bar{4}3m$, which, based on our literature search, has not yet been reported. 
Independently, extensive structure searches performed using AIRSS employing ephemeral data-derived potentials (EDDPs)~\cite{Pickard2022} with `shaken' \textit{fcc} supercells and unit cells of random structures with symmetry operations ranging from 2 to 48, identified the same $P\bar{4}3m$ structure as one of the energetically most favorable, along with a few competitive variants exhibiting similar displacement features, but lower symmetries --- such as tetragonal $P4_2/mcm$ and $P4/nmm$~\footnote{See SI Tables S2 and S3~\cite{SI}. Further analysis shows that $P4/nmm$ and $P4_2/mcm$ are dynamically unstable within both the harmonic approximation and SSCHA, further supporting $P\bar{4}3m$ as the most physically plausible structure among those considered.}. In addition, MD~\cite{PLIMPTON19951}, when averaged over representative snapshots, also converged to the same structure. The consistent identification of the $P\bar{4}3m$ phase by three independent and conceptually distinct methods --- SSCHA, MD, and AIRSS --- provides compelling evidence that this structure represents the true quantum-mechanical ground state of stoichiometric cubic NbN.

The full structural details are provided in the SI~\cite{SI}, Fig.~\ref{fig:comp_phases}(a) provides a comparison between this simple cubic phase and the face-centered cubic $\delta$ phase, illustrating the structural distortions that result from free energy minimization when quantum ionic motion and anharmonicity are considered. Notably, the free energy of the $P\bar{4}3m$ phase is 65\,meV/atom lower than that of the $\delta$ phase --- a significant difference. For a detailed comparison of the free energies among the most relevant 1:1 stoichiometric phases, see Supplemental Tab.~1~\cite{SI}. 

The anharmonic phonon dispersion of $P\bar{4}3m$-NbN, determined using SSCHA, is presented in Fig.~\ref{fig:comp_phonons}(e) as solid blue lines, with the colored background showing the corresponding information obtained from MD-SED. While the overall dispersion resembles that of $\delta$-NbN calculated within the harmonic approximation using large smearing parameters [Fig.~\ref{fig:comp_phonons}(c)], several key differences are evident. The reduced structural symmetry in the $P\bar{4}3m$ phase lifts the high degeneracy observed in the $\delta$ phase, and we observe a phonon hardening for the mode around 6--8\,meV as well as for the high-frequency optical branches.

We also note that harmonic phonon calculations for this structure show imaginary modes only in a narrow region of the Brillouin zone near $\Gamma$, as illustrated in Supplemental Fig.~S4. This localized instability strongly suggests that quantum ionic motion --- specifically zero-point fluctuations --- is the main driving force behind the stabilization of the phase, rather than higher-order anharmonic effects, which typically influence broader regions in reciprocal space.
 
To verify our findings and gain further insight, we performed equilibrium MD simulations and evaluated the vibrational properties using the MD-SED approach. This method fully and implicitly accounts for ionic motion and anharmonicity, making it one of the most rigorous and state-of-the-art techniques for determining vibrational properties~\cite{honarvar2016spectral}, second only to an exact description of quantum ions within \textit{ab initio} path-integral MD~\cite{Marx-1996,ceriotti2018,parrinello2019,piquemal2023}. Starting from the $\delta$ phase, the MD-SED analysis similarly shows that the high-symmetry structure is not maintained, with the ground state adopting a simple cubic symmetry. The phonon spectral function obtained from MD-SED, displayed as the background in Fig.~\ref{fig:comp_phonons}(e), agrees remarkably well with the phonon dispersion calculated via SSCHA. The convergence of results from these two complementary approaches lends strong support to the existence of the novel $P\bar{4}3m$-NbN phase.

\emph{Note:} We extended our investigation to the closely related $fcc$ phases of TiN and NbC (details are provided in the SI~\cite{SI}). For both compounds, no transition to an alternative phase was observed, in contrast to the behavior seen in NbN.

\begin{figure*}[t]
    \centering
    \includegraphics[width=1\linewidth]{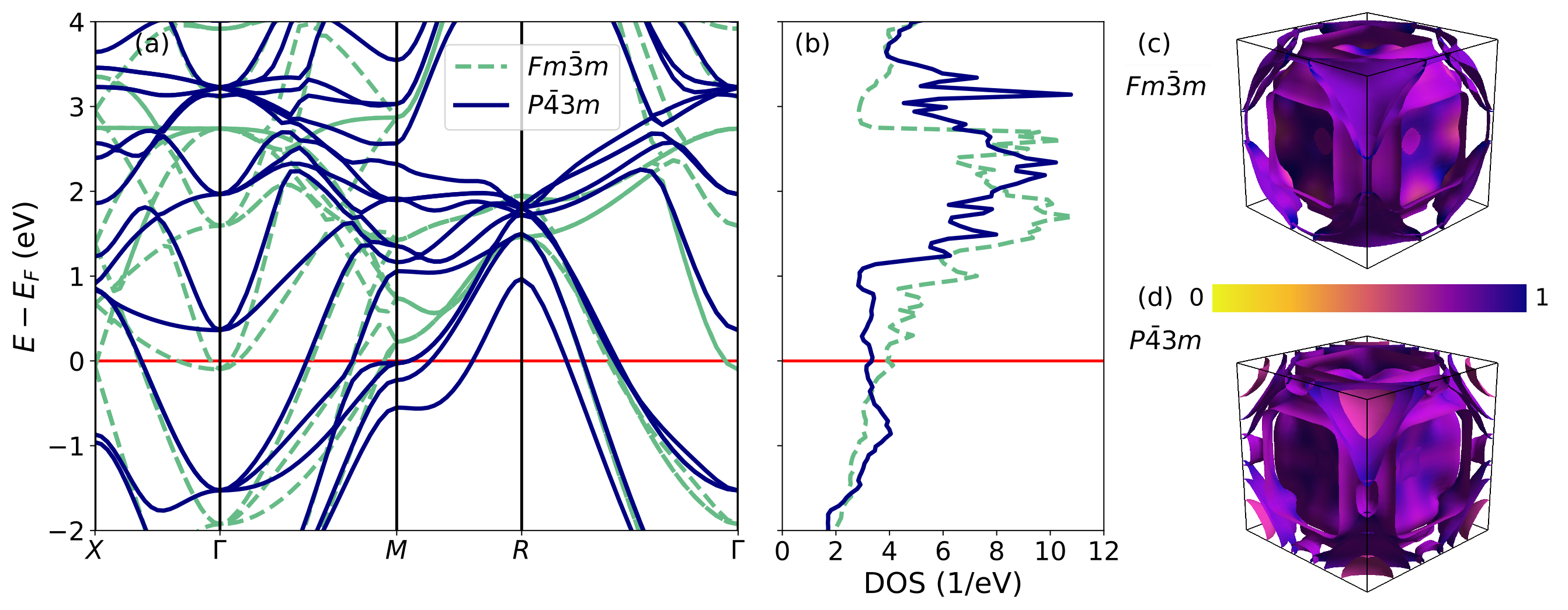}
    \caption{Electronic band structures (a) and DOS (b) for the $Fm\bar{3}m$ (green) and $P\bar{4}3m$ (blue) phases. Fermi surfaces for $Fm\bar{3}$ and $P\bar{4}3m$, visualized with FermiSurfer~\cite{KAWAMURA2019197-fermisurfer}, are shown in (c) and (d), respectively. The color scale indicates the Nb $d$-orbital character, from yellow (0) to blue (1).}
    \label{fig:comp_electrons}
\end{figure*}

In Fig.~\ref{fig:comp_electrons}, we compare the electronic band structures and density of states (DOS) for the $Fm\bar{3}m$ (green) and $P\bar{4}3m$ (blue) phases. Both retain a metallic character, with states near the Fermi level ($E_\mathrm{F}$) predominantly derived from Nb $d$ orbitals. However, the reduced symmetry of the $P\bar{4}3m$ phase lifts several degeneracies evident in the $Fm\bar{3}m$ band structure. In particular, at the $X$ point, bands that cross the Fermi level in $Fm\bar{3}m$ are split and pushed both above and below $E_\mathrm{F}$, effectively removing them from the Fermi surface in $P\bar{4}3m$. Likewise, at $\Gamma$, these bands are pushed up in energy, which further lowers the DOS at $E_\mathrm{F}$. Indeed, as shown in the DOS panel on the right, the blue curve (corresponding to $P\bar{4}3m$) dips below the green curve (corresponding to $Fm\bar{3}m$) near $E_\mathrm{F}$. Such modifications of the electronic states around the Fermi level have a direct impact on the electron-phonon coupling and, as we will show below, play a key role in determining the superconducting properties of NbN.

\begin{figure}[t]
    \centering
    \includegraphics[width=1\linewidth]{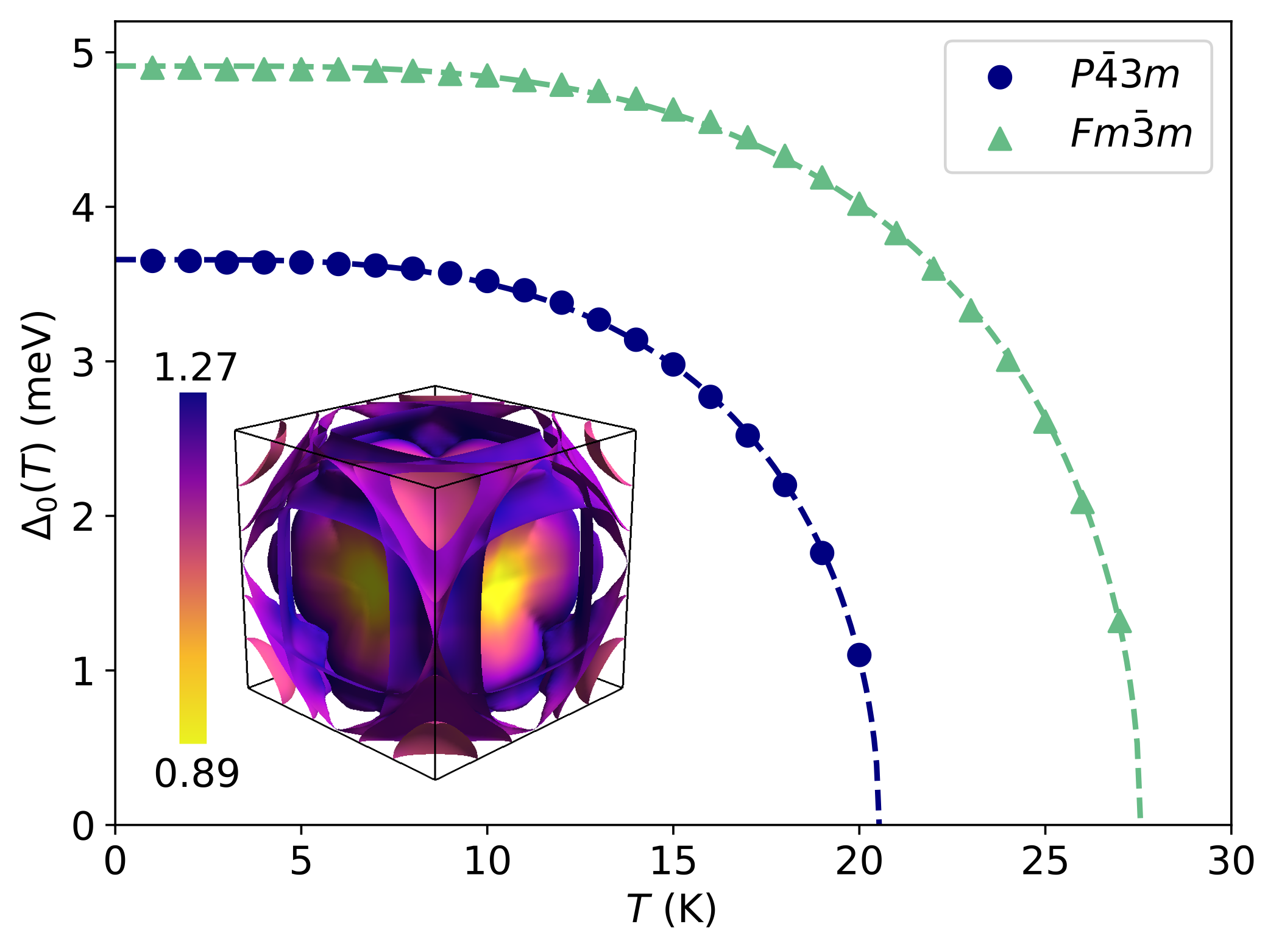}
    \caption{Temperature dependence of the superconducting gap $\Delta_0$ from isotropic ME calculations~\cite{kogler2025isoME}. Results for the P$\bar{4}3m$ and Fm$\bar{3}m$ phases are shown as dots and triangles, respectively. Dashed lines are guides to the eye. The inset shows the P$\bar{4}3m$ Fermi surface, visualized with FermiSurfer~\cite{KAWAMURA2019197-fermisurfer} and colored by $\lambda_\mathbf{k}$ from an anisotropic ME calculation~\cite{lucrezi_full-bandwidth_2024,lee_electronphonon_2023}.}
    \label{fig:comp_SC}
\end{figure}

\begin{table}[t]
    \centering
    \caption{Superconducting parameters for the $Fm\bar{3}m$ and $P\bar{4}3m$ phases of NbN, computed using \textsc{IsoME}~\cite{kogler2025isoME} with a screened Coulomb pseudopotential $\mu = 0.21$ from $GW$~\cite{Lambert2013}. Listed are the electron-phonon coupling constant $\lambda$, logarithmic average phonon frequency $\omega_\text{log}$, and superconducting critical temperatures $T_\text{c}^\text{cDOS}$ and $T_\text{c}^\text{vDOS}$, assuming constant and energy-dependent electronic DOS near the Fermi level, respectively.}  \label{tab:tc}
    \setlength{\tabcolsep}{6pt}
    \begin{tabular}{l|ccccccccccc}
    \toprule
          & $\lambda$ & $\omega_\text{log}$ (meV) & $T_\text{c}^\text{cDOS}$ (K) & $T_\text{c}^\text{vDOS}$ (K) \\
         \midrule
         $Fm\bar{3}m$ & 1.2 & 25 & 28 & 27 \\
         $P\bar{4}3m$      & 1.2 & 19 & 20 & 20 \\
         \bottomrule
    \end{tabular}
\end{table}

Using SSCHA phonon dispersions to compute electron-phonon coupling and $GW$~\cite{Lambert2013} to evaluate the Fermi-surface averaged screened Coulomb interaction $\mu$, we determined the superconducting properties of $P\bar{4}3m$-NbN within the full-bandwidth isotropic Migdal-Eliashberg framework using \textsc{IsoME}~\cite{kogler2025isoME}, complemented by fully anisotropic calculations with EPW~\cite{lucrezi_full-bandwidth_2024}. The EPW results provide a detailed momentum-resolved map of the electron-phonon coupling strength $\lambda_{\mathbf{k}}$, confirming that the interaction is nearly isotropic across the Fermi surface. This justifies the use of the isotropic approximation in \textsc{IsoME}, which we then employ to compute the superconducting gap $\Delta_0(T)$ and obtain a critical temperature of $T_{\text{c}} = 20$\,K, as shown in Fig.~\ref{fig:comp_SC}. This result compares much more favorably with the experimentally obtained $T_{\text{c}} \approx 16\,\text{K}$ for near-stoichiometric bulk NbN than calculations that use large electronic smearing to artificially stabilize the $\delta$ phase, which yield $T_{\text{c}} = 27$\,K. Considering the experimental trend of increased \Tc{} with decreasing vacancy concentration, our results suggest that achieving stoichiometric 1:1 cubic NbN could lead to even higher superconducting transition temperatures.

\emph{Note:} To assess the importance of the electronic dispersion near the Fermi level, we computed superconducting critical temperatures employing the infinite-bandwidth-constant-DOS (cDOS) and the finite-bandwidth-energy-dependent-DOS (vDOS) approximations within the isotropic ME framework (for more details see Ref.~\cite{kogler2025isoME}). As stated in Tab.~\ref{tab:tc}, both $Fm\bar{3}m$ and $P\bar{4}3m$ phases yield nearly identical $T_\text{c}$ values across the two treatments.

The fact that the $P\bar{4}3m$ phase is dynamically stable, thermodynamically more favorable, and shows superconducting properties that align well with experimental findings raises the question of whether this phase has been observed experimentally. 
An extensive literature search~\cite{Christensen1977, Miura2013, Treece1995, Heger1980, Yen-1967, Duwez_1950, Brauer-1961, Brauer-1979, Christensen1976, Christensen1977-TinZrnNbNNbc,Christensen1981, Brauer-1952, Pessall1968, Lengauer1986, Oya-1974, Terao_1965, CHRISTENSEN1978} reveals surprisingly few reports characterizing a high-quality, low vacancy-concentration cubic phase of NbN.
In the few studies that include X-ray or neutron diffraction data, the additional weak diffraction peaks associated with the $P\bar{4}3m$ structure --- as shown in Fig.~\ref{fig:comp_phases}(b) --- are absent~\cite{Brauer-1952, Miura2013, Christensen1977}. However, high-pressure measurements below 6\,GPa reveal faint traces of a diffraction feature around the expected 110 reflection of the $P\bar{4}3m$ phase. In contrast, no signature is observed at the 100 reflection~\cite{Chen-2-2005, Tan2017}.

One plausible explanation for the elusive nature of the $P\bar{4}3m$ structure is the presence of local distortions within the crystal. Random displacements of Nb and N atoms may arise due to vacancies of nitrogen atoms typically reported for $\delta$-NbN, with stoichiometry in the range $x \approx 0.88$–$0.98$. To date, only a single crystal study of $\delta$-NbN has been reported, in which small static distortions of the Nb and N sublattices were observed. However, crystal structure elucidation was based exclusively on the expected diffraction peaks from the $Fm\bar{3}m$ space group~\cite{Heger1980}.

In summary, our comprehensive computational investigation has revealed a previously unreported, dynamically stable simple cubic phase of NbN (space group $P\bar{4}3m$), which is thermodynamically more favorable than the conventional $\delta$-NbN phase. By employing both the SSCHA and MD-SED methodologies, we have shown that quantum anharmonic effects --- in particular zero-point motion and ionic potential non-parabolicity --- stabilize this novel phase, leading to a reduced electron-phonon coupling compared to that of the idealized stoichiometric $\delta$ phase. Consequently, the superconducting transition temperature ($T_{\text{c}} = 20$\,K) calculated for the $P\bar{4}3m$ phase aligns much more closely with experimental observations for near-stoichiometric $\delta$-NbN than previous estimates based on large electronic smearing approximations.

These findings challenge the conventional view that vacancies are necessary to stabilize cubic NbN and indicate that synthesizing the vacancy-free, distorted cubic 1:1 phase could offer a promising pathway to enhancing superconductivity in NbN, potentially reaching even higher critical temperatures. We encourage further experimental investigations --- especially using high-resolution diffraction techniques on single NbN crystals or well-controlled powders --- to validate the presence of the $P\bar{4}3m$ phase and fully assess its potential in superconducting applications. Future studies might also explore the impact of strain and chemical substitution on the phase stability and superconducting performance of NbN, as well as similarly explore the effects of quantum anharmonicity on the stability of other superconducting materials.

\section*{Methods}
\subsection{DFT and DFPT calculations}
Electronic structure calculations are performed using density functional theory (DFT), as implemented in the Quantum ESPRESSO package \cite{QE-2009, QE-2017, QE-2020} with the PBE exchange-correlation functional \cite{GGA_perdew_1996PhysRevLett.77.3865} and scalar-relativistic optimized norm-conserving Vanderbilt pseudopotentials \cite{vanbilt_pseudo_hamann_2013_PhysRevB.88.085117}. A plane-wave basis set with a cutoff energy of 120\,Ry was employed to expand the Kohn-Sham orbitals. The Brillouin zone was sampled using a $18 \times 18 \times 18$ $\mathbf{k}$-grid. A Gaussian spreading of 0.01\,Ry was applied for all structures, except for the \textit{large smearing} calculations, where the smearing was set to 0.3\,Ry.
Structural relaxations were converged to 10$^{-7}$\,Ry in total energy and 10$^{-6}$\,Ry/a$_0$ in forces. DFPT calculations were performed on a $4 \times 4 \times 4$ $\mathbf{q}$-grid with a self-consistency threshold of 10$^{-14}$\,Ry or lower.

\subsection*{AIRSS}

We used AIRSS~\cite{pickard2006high,AIRSS}, accelerated by EDDPs~\cite{Pickard2022}, to search for novel NbN structures. The search was constrained to cells with 2--48 symmetry operations and 64 atoms with a 1:1 Nb:N ratio. Approximately 50,000 structures were generated and relaxed using EDDPs. The final structures were then filtered by an energy window of 100 meV per atom, and then by crystal system (removing hexagonal structures). The potential was trained on a dataset of single-point energies of approximately 12,000 structures, across a stoichiometry range of Nb$_x$N$_y$, x,y=0-4. In addition, `marker' structures of known low energy Nb$_x$N$_y$ structures were `shaken' and added to the dataset, including the $P\bar{4}3m$ structure. Single-point energy calculations were performed in \textsc{CASTEP}~\cite{Clark2005}, using the PBE exchange-correlation functional\cite{GGA_perdew_1996PhysRevLett.77.3865} with on-the-fly generated C19 pseudopotentials, a plane-wave cutoff of 600\,eV, and a $\mathbf{k}$-point spacing of $0.04 \times 2\pi/\text{\AA}$. The EDDP used a cut-off radius of $5.5$\,\text{\AA}, a neural network of 2 hidden layers each with 20 nodes and a feature vector comprised of 16 and 4 polynomials for two- and three-body interactions respectively. We also performed relaxations of `shaken' \emph{fcc} cells with 2-6 formula units of NbN, directly within DFT. The best candidates from both searches were further refined using DFT calculations with \textsc{CASTEP} (See SI Tables S2 and S3~\cite{SI}). These used a slightly higher plane-wave cutoff of 800\,eV, and a smaller $\mathbf{k}$-point spacing of $0.02 \times 2\pi/\text{\AA}$. 

\subsection*{MTP}
The potential was trained using the MLIP package~\cite{Novikov_2021}, at level 26, including eight radial basis functions and a cutoff $R_\text{cut}=5.0$\,\AA. The training set consisted of 100 structures, including 50 layered system structures and 50 bulk structures, randomly chosen out of all individuals of the DFT SSCHA calculations for $\delta$-NbN at 300\,K. The set for the layered (bulk) system comprised 25 structures in the 2$\times$2$\times$1 (2$\times$2$\times$2) supercell and 25 structures in the 3$\times$3$\times$1 (3$\times$3$\times$3) supercell. We validated the potential on distorted structures around $\delta$-NbN, as well as on structures near the $P\bar{4}3m$ phase, yielding an RMSE of 0.6-0.7\,meV/atom for the total energy and 60-80\,meV/\AA{} for the force components. The graphical validation is shown in Supplemental Fig.~S8.

\subsection*{SSCHA}
SSCHA calculations were carried out in constant-volume mode with structural relaxation, using the SSCHA Python package~\cite{Monacelli_2021} and following the workflow described in Refs.~\cite{lucrezi2023quantum,lucrezi2024}. Total energies, forces, and stress tensors for each ensemble configuration were computed using a purpose-trained MTP~\cite{Novikov_2021}, as described in the previous section.

We performed SSCHA simulations using 2$\times$2$\times$2, 3$\times$3$\times$3, and 4$\times$4$\times$4 supercells, considering both fixed and non-fixed symmetry. To reduce stochastic noise, the final ensemble for each supercell included 10,000 configurations. The ratio between the free energy gradient (with respect to the auxiliary dynamical matrix) and its stochastic error was reduced below $10^{-7}$. The Kong-Liu effective sample size was set to 0.6.
Anharmonic phonon dispersions were extracted from the positional free energy Hessian, evaluated in the bubble approximation (excluding the fourth-order term). The phonon spectra obtained from the final two ensembles differ by less than 1\,meV.

For the $Fm\bar{3}m$ phase, calculations were performed on supercells of the primitive $fcc$ unit cell, where all atoms occupy high-symmetry Wyckoff positions. Symmetry detection was enabled during the SSCHA minimization.

For the $P\bar{4}3m$ phase, supercells were constructed from the conventional (simple cubic) unit cell. Symmetry detection was disabled both during the preparatory DFPT calculations and the SSCHA minimization, allowing all atomic positions to relax freely.

\subsection*{MD simulations and SED calculations}
The MD-SED method allows us to obtain the phonon band structure directly from equilibrium MD simulations where anharmonic effects are fully accounted for. The method comprises applying a Fourier transform to the MD velocity field to obtain a frequency-versus-wavenumber mapping of the energy density along the wave-vector direction of interest. In this work, the SED approach used requires knowledge of only the crystal unit-cell structure and does not require any prior knowledge of the phonon mode eigenvectors. The SED expression is given by~\cite{larkin2014comparison,thomas2010predicting}
\begin{equation}
{{{\Phi'}}\left( \pmb{\kappa}, \omega \right)} = 
\displaystyle\sum\limits_{\alpha}^{3}
\displaystyle\sum\limits_{b}^{n}
\begin{vmatrix}
{\mu_{0}\displaystyle\sum\limits_{l}^{N}{ \displaystyle\int\limits_{0}^{\tau_{0}}{\dot{{u}}_{\alpha}
\left (\scriptsize{\!\!\!\!
\begin{array}{l l} 
\begin{array}{l} l\\b
\end{array} \!\!\!\!\!\!
 &;~ t
\!\!\!\!\end{array}}
\right )
e^{{\textrm{i}}\left[\pmb{\kappa}\cdot\pmb{r}_{0}
\left (\!\!
\scriptsize{\begin{array}{l} l\\0
\end{array} \!\!}
\right )-wt\right]} \textrm{d} t}} }
\end{vmatrix}
^{2} , \nonumber
\label{eq:SED_Jonas}
\end{equation}
\noindent where ${\dot{{u}}}_{\alpha}$ is the $\alpha$-component of the velocity of the $b$th atom in the $l$th unit cell at time $t$, $\mu_{0}=m_{b}/(4\pi \tau_{0} N)$, $\tau_{0}$ is the total simulation time, and ${\pmb{r}}_{0}$ is the equilibrium position vector of the $l$th unit cell. Our MD model is based on a cubic NbN unit cell with a side length of $a=4.482262\,\text{\AA}$. There are $n=8$ atoms per unit cell and a supercell comprising of $N=N_{x}\times N_{y}\times N_{z} = 30 \times 30 \times 30$ unit cells forming the total simulated computational domain. This corresponds to a wave-vector resolution of ${\Delta}{\kappa} = 0.033\,(2\pi/a)$ along each principal direction in the wavenumber space. We note that in Eq.~(\ref{eq:SED_Jonas}), the phonon frequencies can be obtained only for the set of allowed wave vectors as determined by the crystal structure. The MD simulations are performed using the LAMMPS software \cite{PLIMPTON19951}. The system is first equilibrated under the $NPT$ ensemble at 16\,K for 0.2\,ns. After equilibration, the ensemble is switched to $NVE$, and the system is simulated for $2^{20}$ time steps with a time interval of ${\Delta}{t} = 0.5$\,fs. Eq.~(\ref{eq:SED_Jonas}) is evaluated using the velocity trajectories extracted every $2^{5}$ steps.

\subsection*{Isotropic Migdal-Eliashberg Calculations}
Isotropic superconducting properties were determined using IsoME~\cite{kogler2025isoME}. A value of $\mu=0.21$ was applied, obtained via the $GW$ approximation within SternheimerGW~\cite{Lambert2013}. The Matsubara frequency cutoff, \textit{omega\_c}, was set to 7,000\,meV. In variable DOS computations, the chemical potential was updated continuously. An energy cutoff (\textit{encut}) of 5,000\,meV was applied to all quantities, except for the energy shift and $\mu$ update calculations, where a lower cutoff (\textit{shiftcut}) of 2,000\,meV was used.

\subsection*{Anisotropic Migdal-Eliashberg Calculations}
Anisotropic ME calculations were carried out using the \textsc{EPW} code~\cite{lee_electronphonon_2023, lucrezi_full-bandwidth_2024}. The construction of maximally localized Wannier functions was initialized via the SCDM algorithm. Electron-phonon matrix elements were computed on a coarse Brillouin zone mesh of $8\times8\times8$ $\mathbf{k}$-points and $4\times4\times4$ $\mathbf{q}$-points, and subsequently interpolated onto a dense grid of $24\times 24 \times 24$ $\mathbf{k}$-points and $12\times12\times12$ $\mathbf{q}$-points for accurate evaluation of the momentum-resolved superconducting properties.

\section*{Acknowledgments}
We thank Stuart A. Wolf for many interesting and insightful discussions on superconducting alloys and NbN in particular.

This work was funded in part by the Defense Sciences Office (DSO) of the Defense Advanced Research Projects Agency (DARPA) and Intellectual Ventures Property Holdings.
Computations were performed on the lCluster at TU Graz, the Vienna Scientific Cluster (VSC5), and the Ohio Supercomputing Center (OSC).
R.L. acknowledges the Carl Tryggers Stiftelse för Vetenskaplig Forskning (CTS 23: 2934).

\section*{Author contributions}
E.K., F.J., R.L., and C.H. performed the MTP and SSCHA calculations. E.K., R.L., and M.R.S. conducted DF(P)T computations. C.N.T. performed the MD simulations and SED calculations. P.I.C.C. and C.J.P. performed the AIRSS calculations. E.K. and C.H. calculated superconducting properties. B.K. and R.R. carried out a literature search and contributed to interpreting the results.
M.N.J., R.P.P, M.I.H. and C.H. supervised the project.
All authors contributed to discussions and revised the manuscript. 

\section*{Data Availability}
The authors confirm that the data supporting the findings of this study are available within this article or its Supplementary Information~\cite{SI}. 

\bibliographystyle{apsrev4-2}
\bibliography{literature}

\clearpage
\newpage
\begin{figure*}
\includegraphics[page = 1, width=\linewidth]{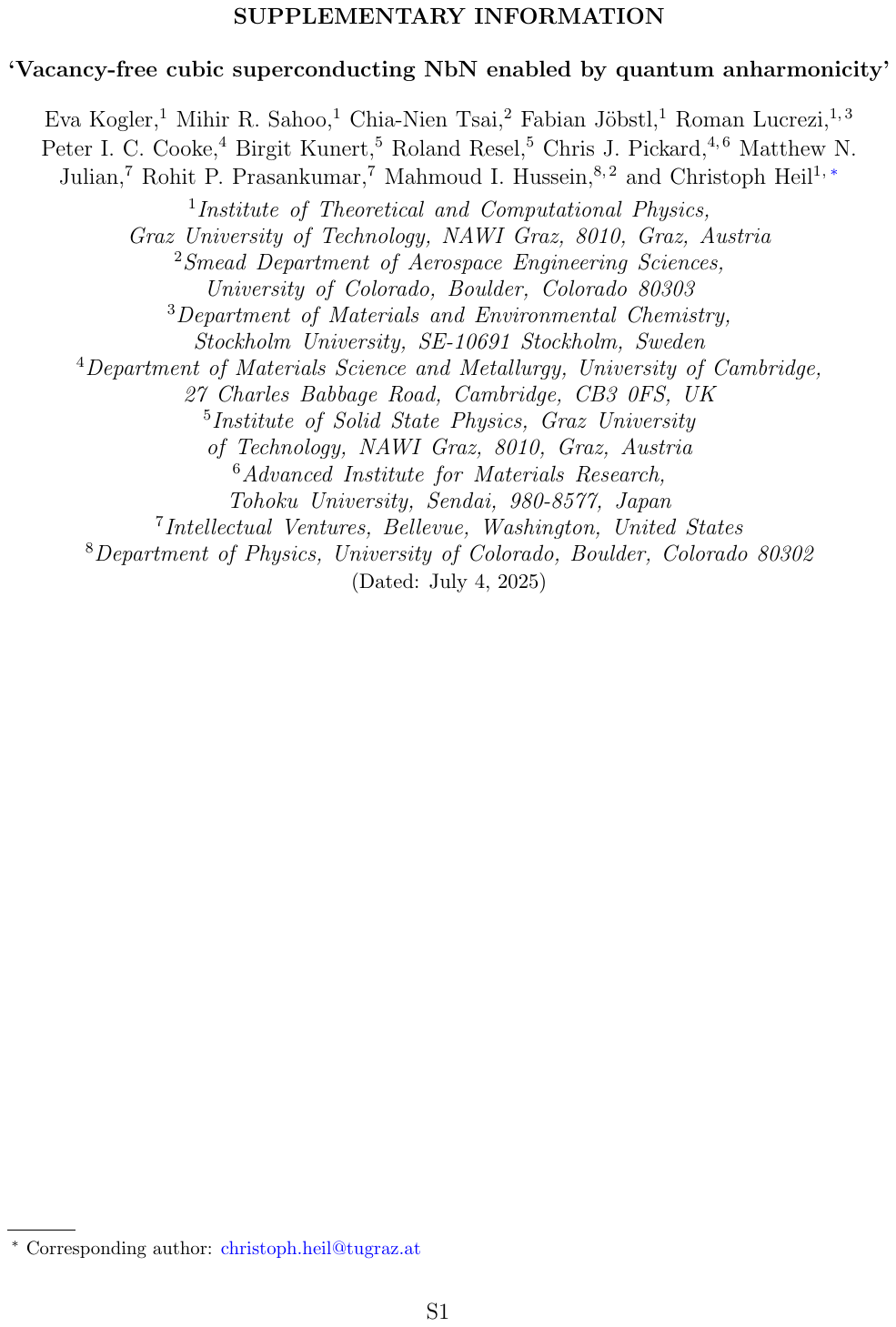}
\end{figure*}
\begin{figure*}
\includegraphics[page = 2, width=\linewidth]{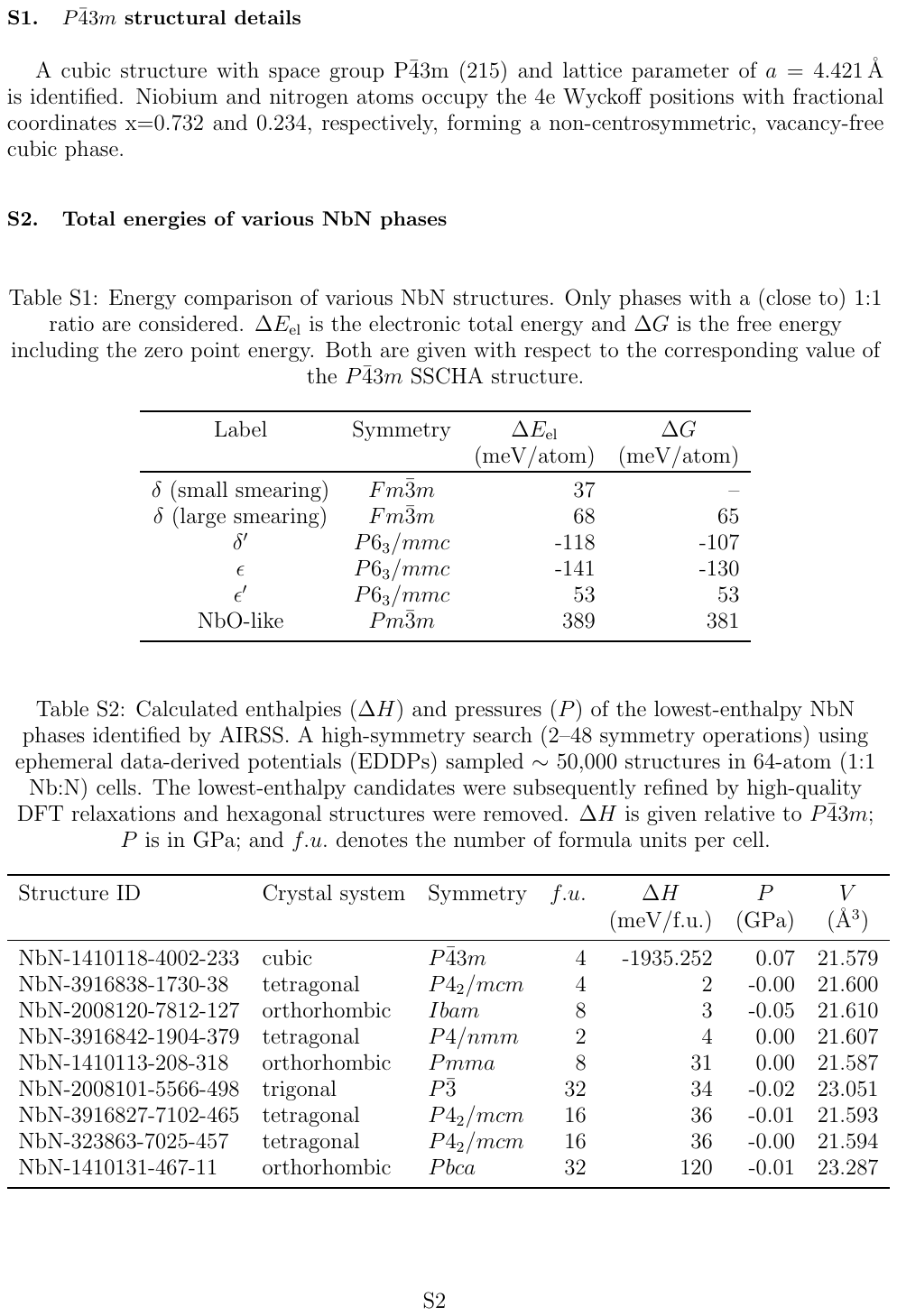}
\end{figure*}
\begin{figure*}
\includegraphics[page = 3, width=\linewidth]{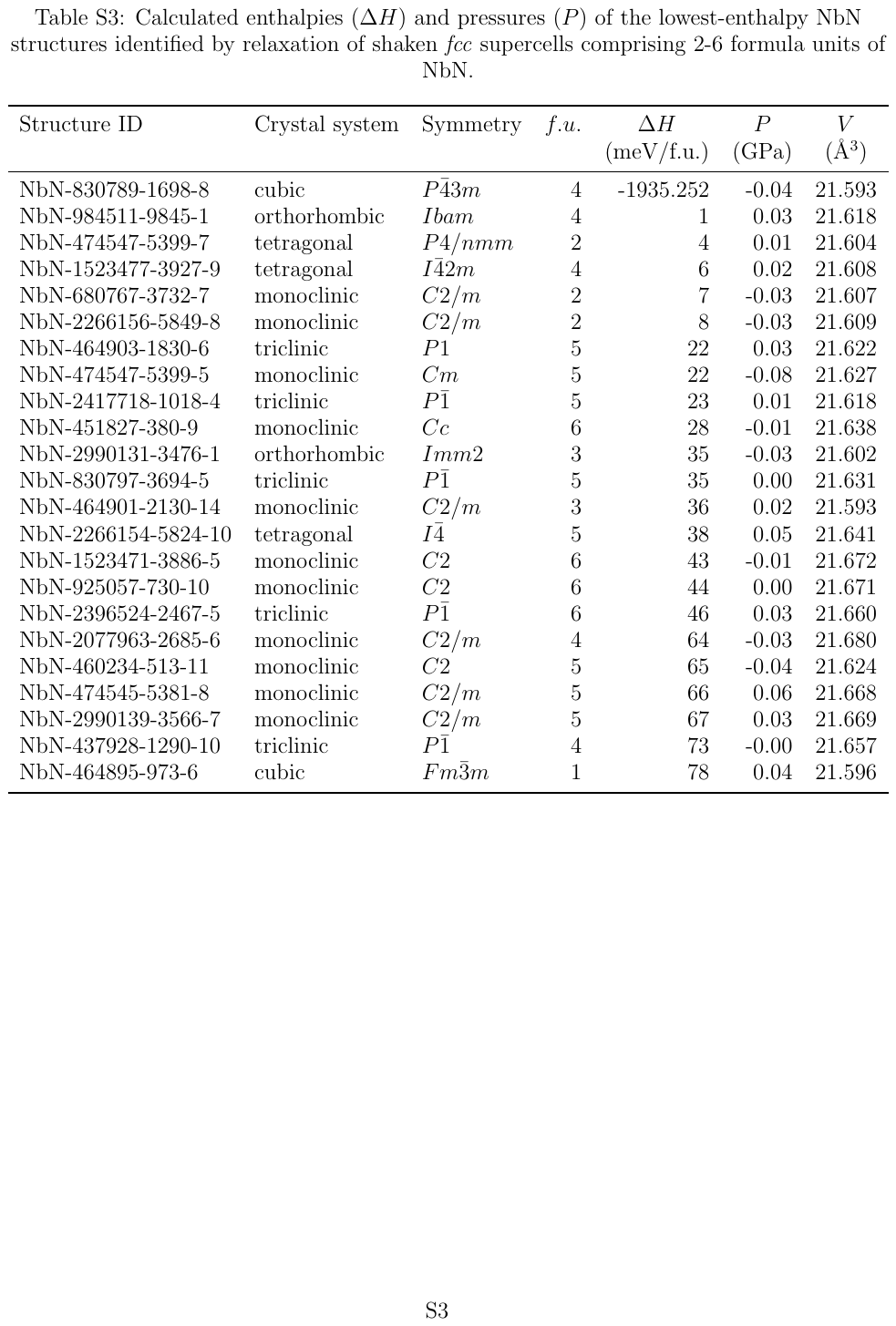}
\end{figure*}
\begin{figure*}
\includegraphics[page = 4, width=\linewidth]{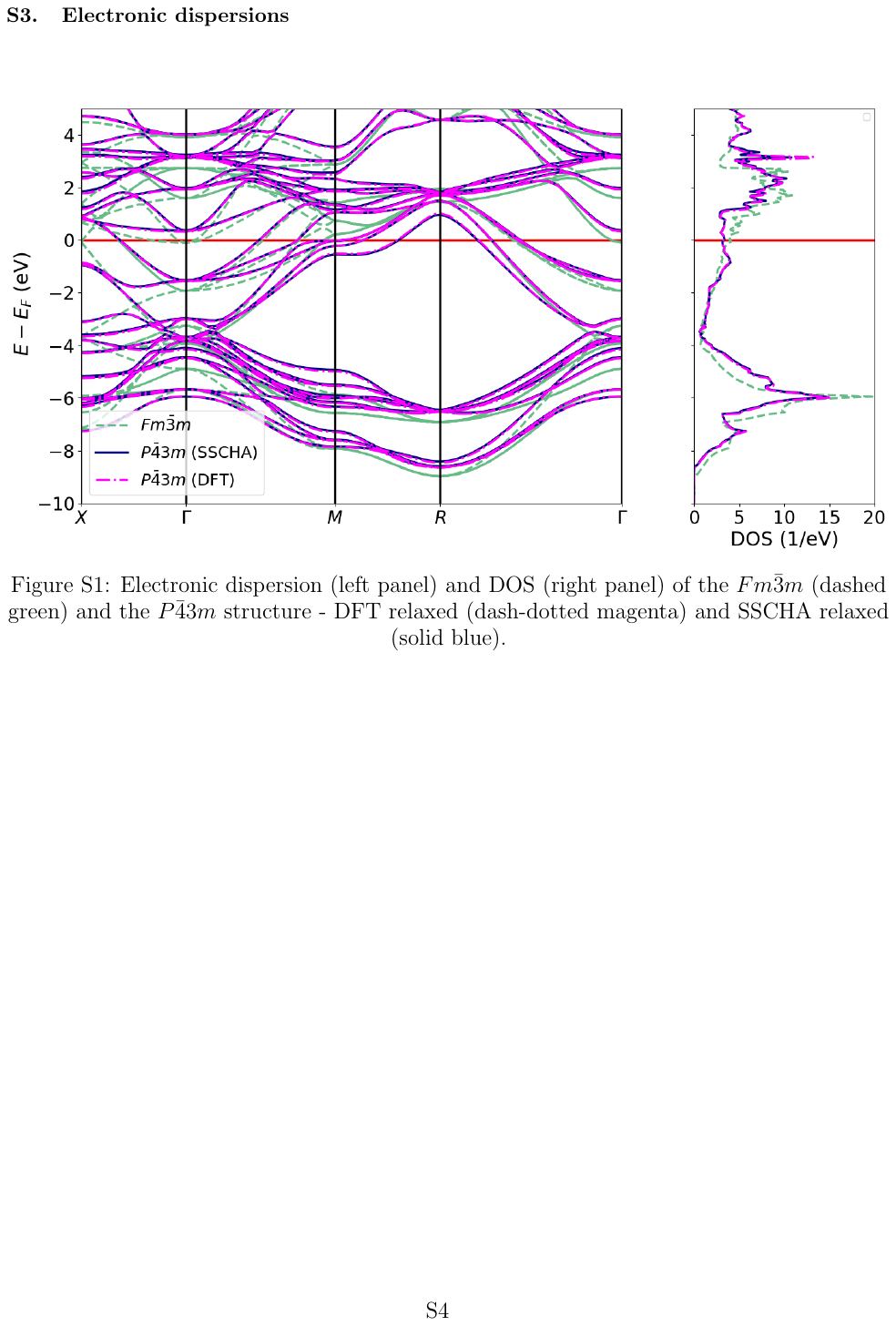}
\end{figure*}
\begin{figure*}
\includegraphics[page = 5, width=\linewidth]{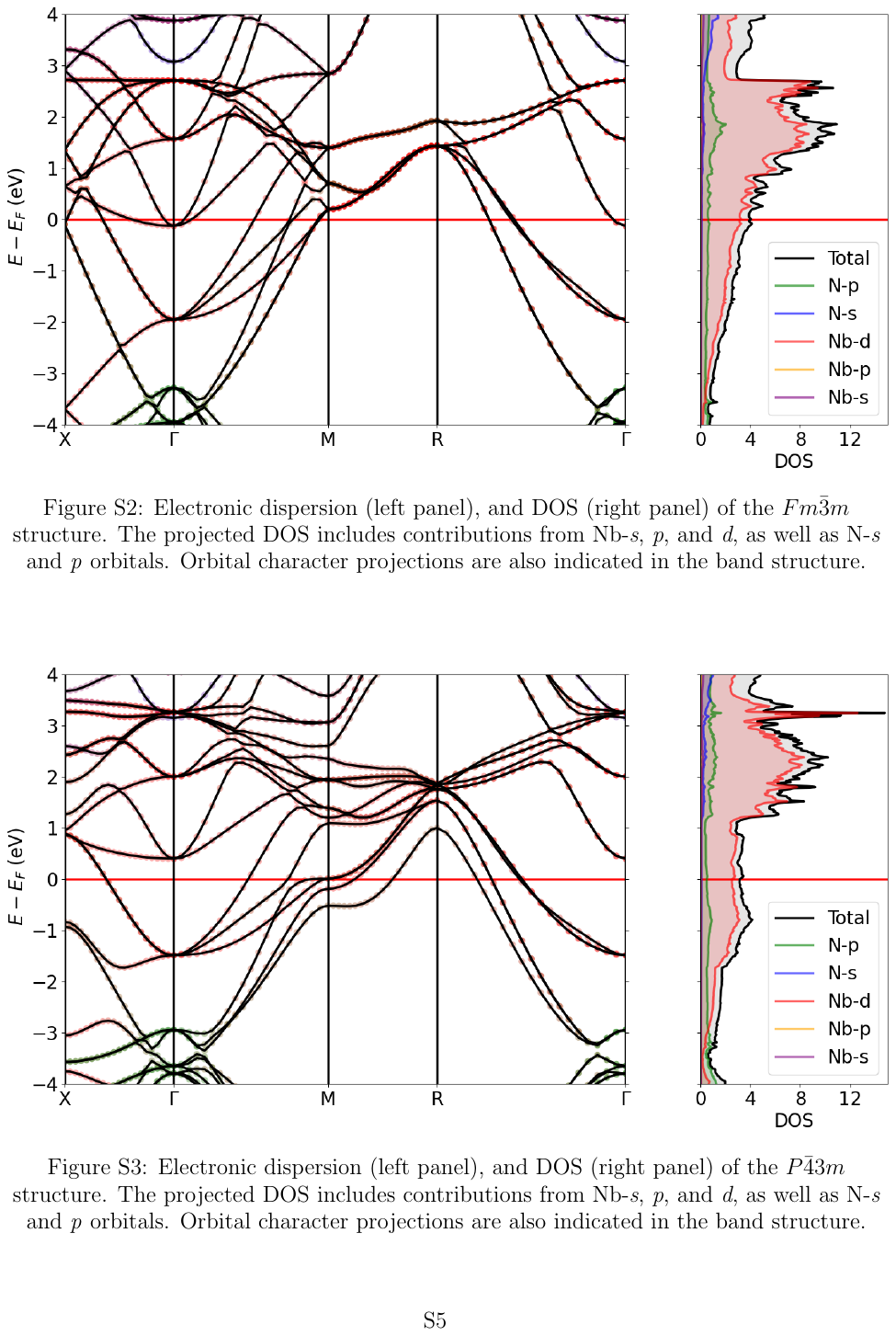}
\end{figure*}
\begin{figure*}
\includegraphics[page = 6, width=\linewidth]{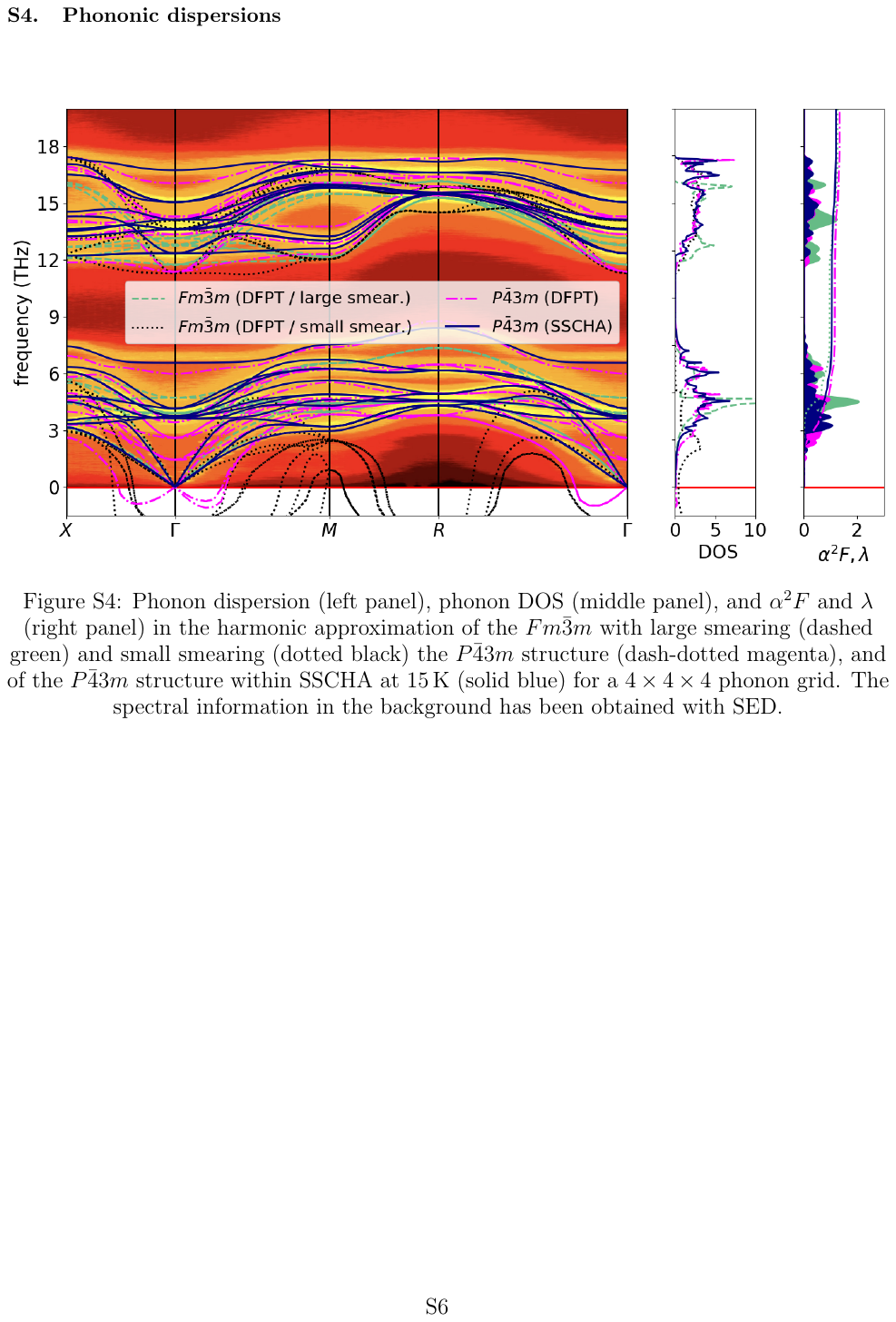}
\end{figure*}
\begin{figure*}
\includegraphics[page = 7, width=\linewidth]{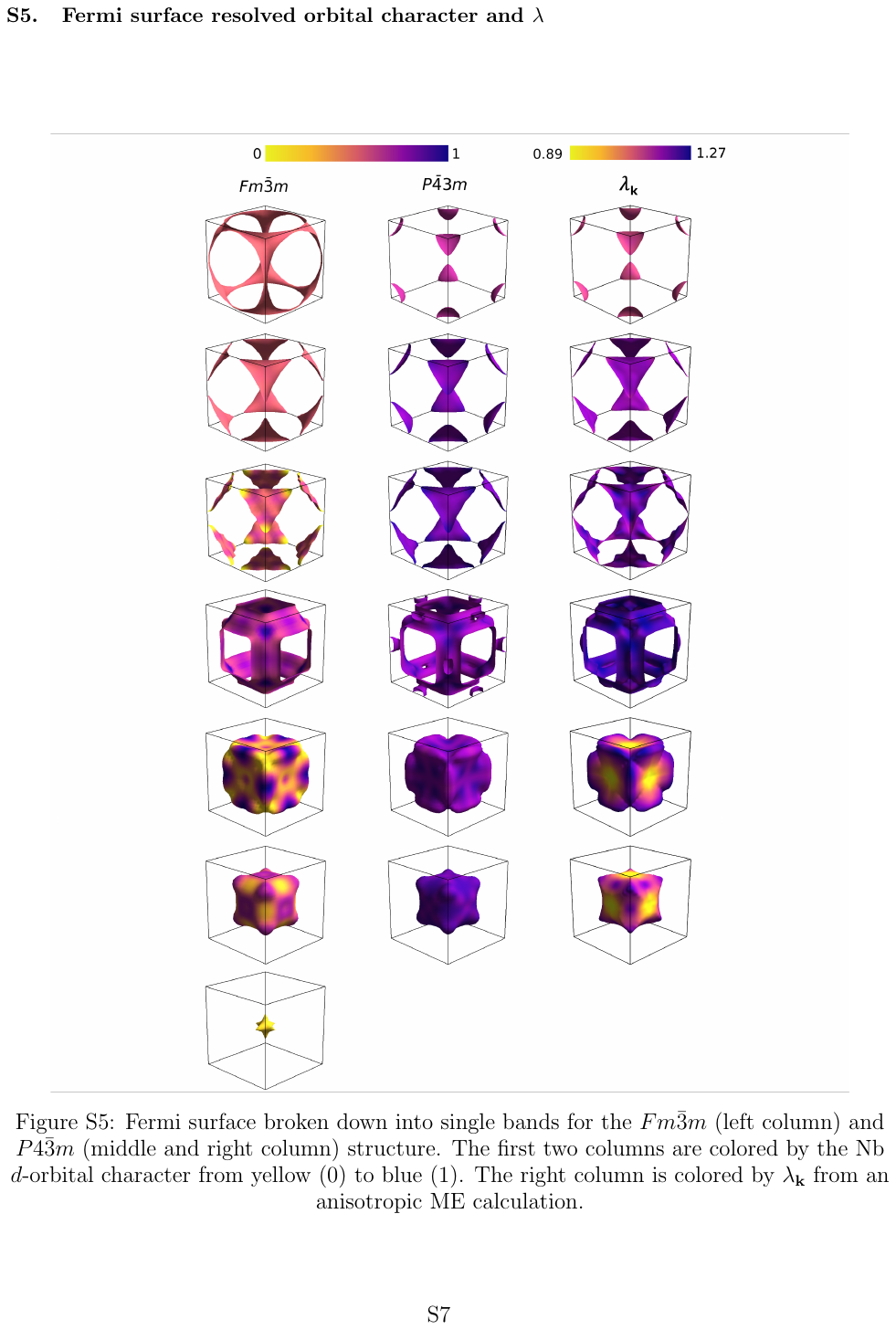}
\end{figure*}
\begin{figure*}
\includegraphics[page = 8, width=\linewidth]{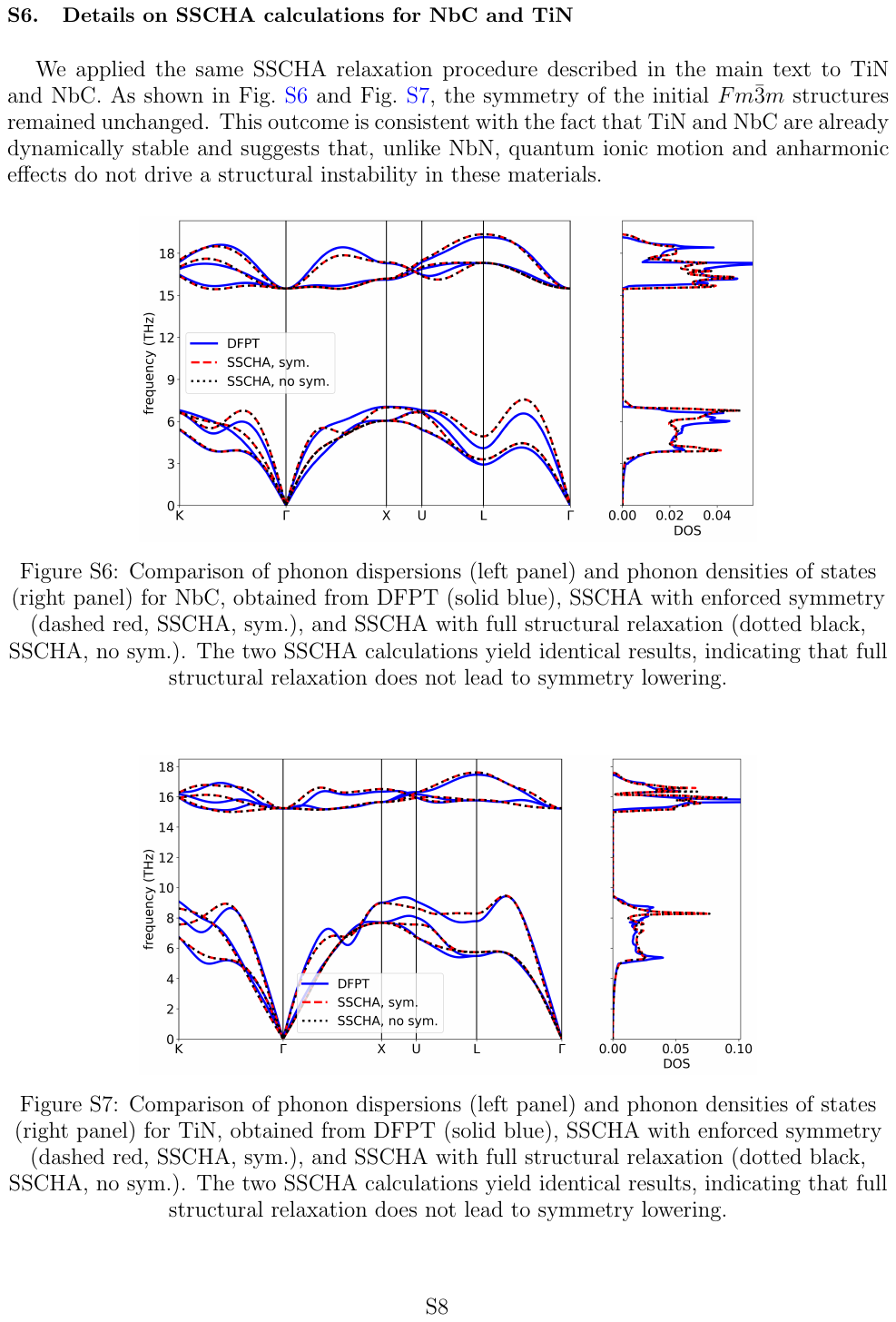}
\end{figure*}
\begin{figure*}
\includegraphics[page = 9, width=\linewidth]{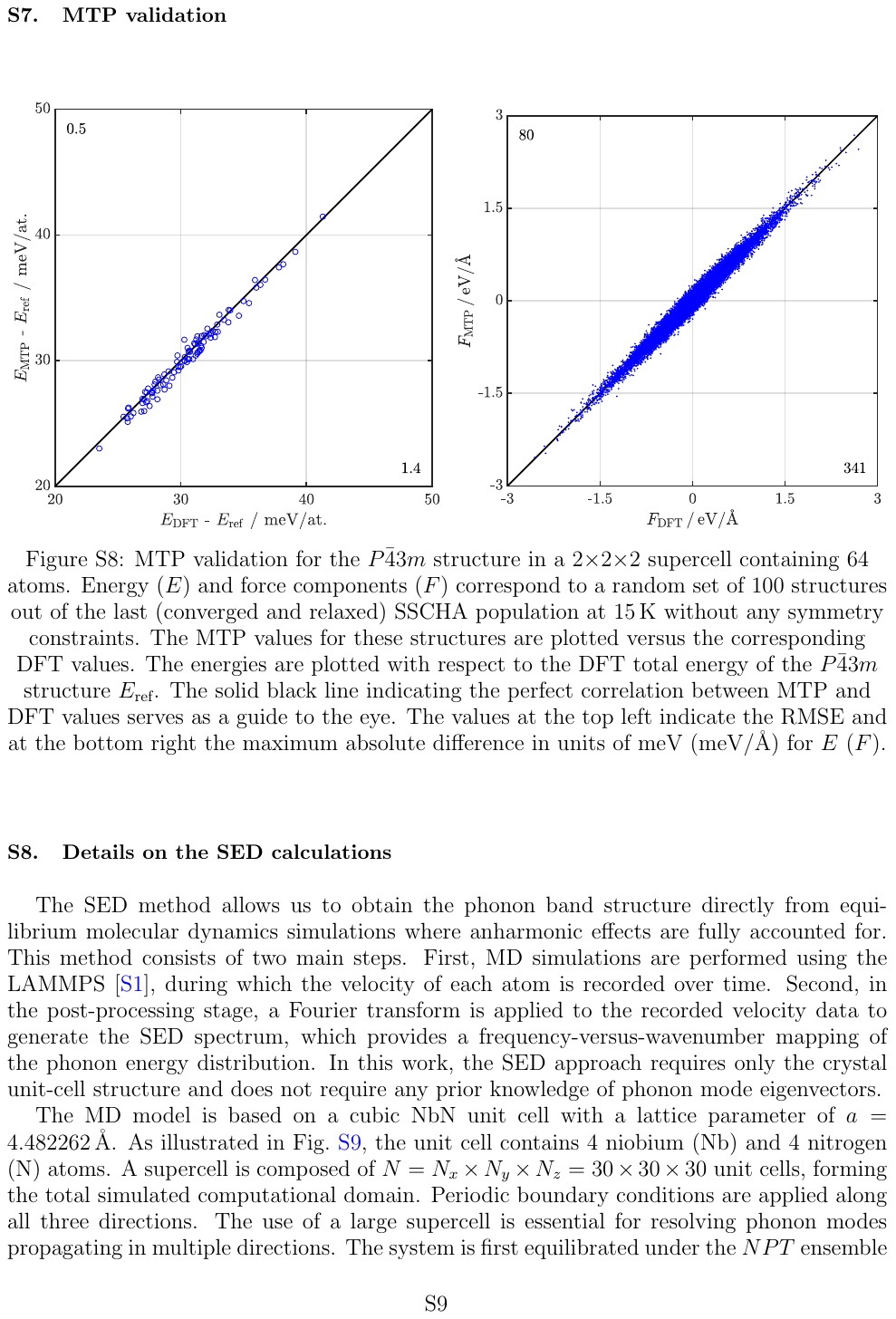}
\end{figure*}
\begin{figure*}
\includegraphics[page = 10, width=\linewidth]{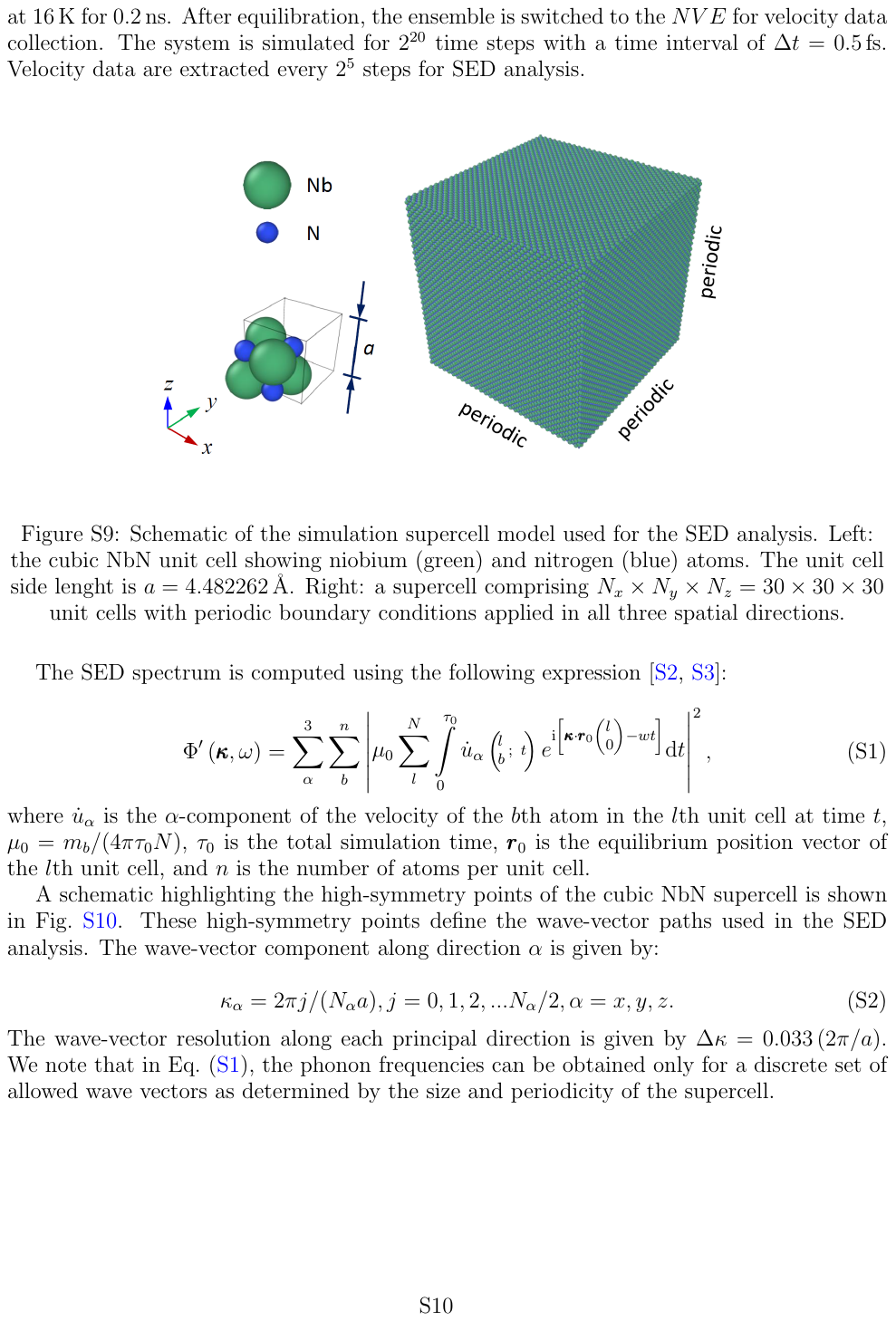}
\end{figure*}
\begin{figure*}
\includegraphics[page = 11, width=\linewidth]{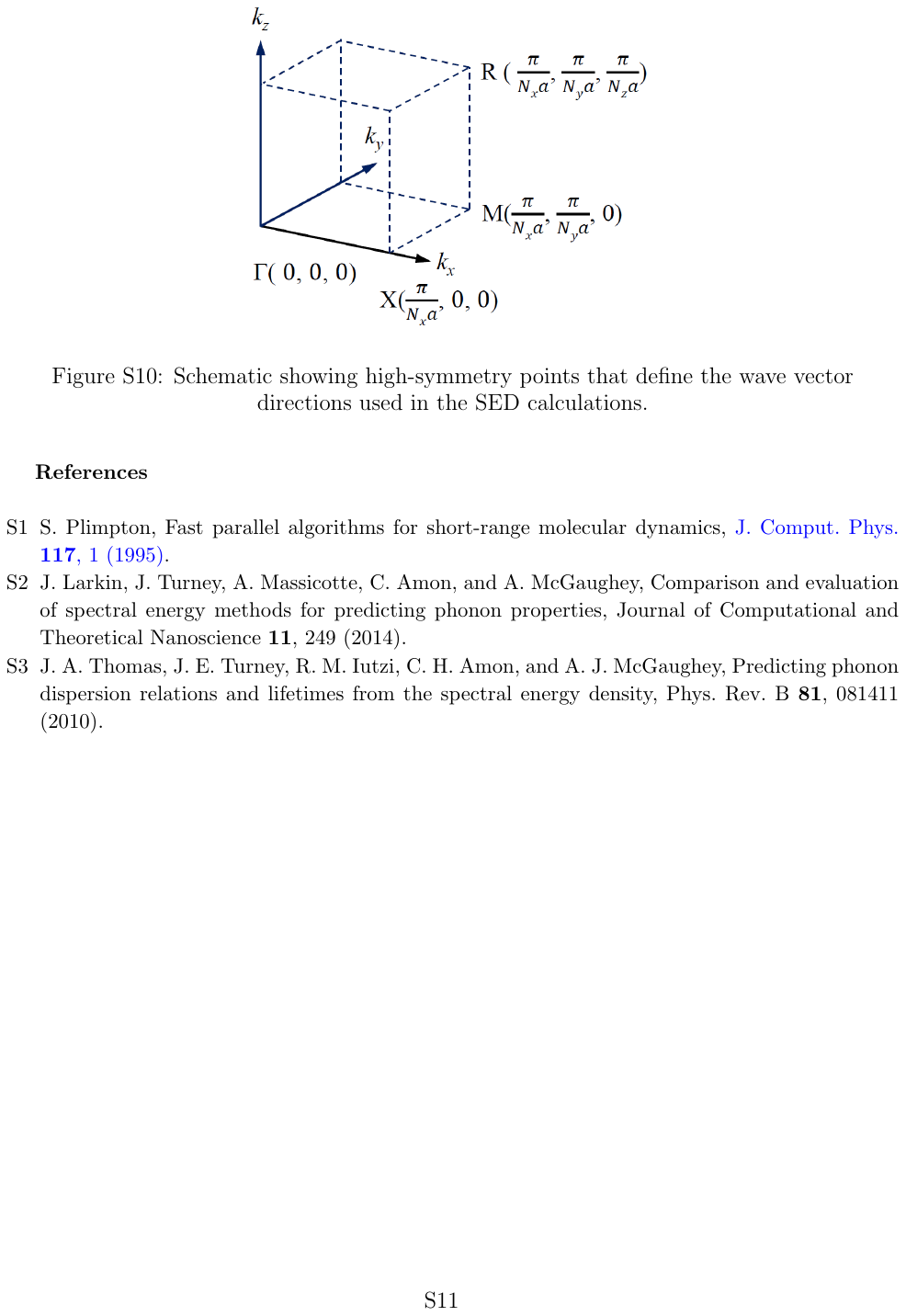}
\end{figure*}

\end{document}